\documentclass[pdflatex,sn-mathphys-num]{sn-jnl}

\usepackage{graphicx}%
\usepackage{multirow}%
\usepackage{amsmath,amssymb,amsfonts}%
\usepackage{amsthm}%
\usepackage{mathrsfs}%
\usepackage[title]{appendix}%
\usepackage{xcolor}%
\usepackage{textcomp}%
\usepackage{manyfoot}%
\usepackage{booktabs}%
\usepackage{algorithm}%
\usepackage{algorithmicx}%
\usepackage{algpseudocode}%
\usepackage{listings}%

\PassOptionsToPackage{unicode}{hyperref}
\PassOptionsToPackage{hyphens}{url}
\def\mybibitem{%
  \vskip\baselineskip
  \noindent
  \hangindent=1.5em
}
\begin{document}

\title{ELECTRIC FIELDS AND WAVES IN THE VENUS NIGHTSIDE MAGNETOSPHERE}


\author*[1,2,]{\fnm{Forrest} \sur{Mozer}}\email{FM@gmail.com}

\author[1]{\fnm{A. V.} \sur{Agapitov}}\email{AVA@gmail.com}

\author[1,2]{\fnm{S. D.} \sur{Bale}}\email{SDB@gmail.com}
\author[1]{\fnm{J. W.} \sur{Bonnell}}\email{JWB@gmail.com}
\author[1]{\fnm{M.} \sur{Pulupa}}\email{MP@gmail.com}
\author[1]{\fnm{T.} \sur{Quinn}}\email{TQ@gmail.com}
\author[3]{\fnm{A.} \sur{Voschchepynets}}\email{AV@gmail.com}

\affil[1]{\orgdiv{Space Sciences Laboratory}, \orgname{University of California, Berkeley},
\orgaddress{\city{Berkeley}, \postcode{94720}, \state{California}, \country{USA}}}

\affil[2]{\orgdiv{Department of Physics}, \orgname{University of California, Berkeley}, \orgaddress{\city{Berkeley}, \postcode{94720}, \state{California}, \country{USA}}}

\affil[3]{\orgdiv{Department of Physics}, \orgname{Uzhhorod National University}, \orgaddress{\city{Uzhhorod}, \country{Ukraine}}}


\abstract{
On November 6, 2024, the Parker Solar Probe flew past Venus to make the
first accurate electric field measurement in the nightside Venusian
magnetosphere. To achieve this result, the electric field antennas were
current biased in a way never before experienced by an electric field
detector. This biasing requirement, that the positive bias
current in the Venus shadow be about equal to the electron thermal
current, is discussed and illustrated. About one minute of useful
electric field data in the eight-minute nightside magnetosphere crossing
was obtained, during which the only feature observed was a few Hz
signal. This result, along with the magnetic field measurements, showed
that there were few if any electromagnetic waves, such as low-frequency
electromagnetic turbulence or whistlers, in the nightside crossing.
Instead, a few Hertz, purely electrostatic signal was found. This
suggests that the interaction of the solar wind with an unmagnetized
body having an ionosphere may be different from that of previously
studied magnetized bodies. In the sunlit flanks, many electromagnetic
wave modes were observed.  These results describe the first step in the proper technique for future measurements of electric fields in shadow.

\break
\break
\textbf{Plain language summary.}
The electric field instrument on the Parker Solar Probe had a bias
current that was varied while the spacecraft was in the Venusian
nightside magnetosphere, such that there were times when the bias
current was appropriate for an accurate measurement of the electric
field. Low-frequency electrostatic turbulence was observed and no
electromagnetic waves, such as whistlers, were seen.
}




\maketitle

\section{Introduction}\label{introduction}

The planet Venus does not have an intrinsic dipole-like magnetic field
{[}Bridge et al, 1967{]}. In the absence of such a field, the
electrodynamic interaction between the Venus ionospheric plasma and the
solar wind creates an induced magnetosphere that may drive a variety of
plasma waves. Waves in magnetized plasmas, such as a turbulent cascade,
MHD waves, whistler waves, etc., may not exist at Venus, in which case
Venus has a plasma environment that has not been well studied.

Electric field instruments have been flown near Venus on the Pioneer
Venus Orbiter {[}Taylor et al, 1979; Scarf et al, 1980{]}, Galileo
{[}Gurnett et al. 1991{]}, and Cassini {[}Gurnett et al. 2001{]} with a
primary aim of determining whether lightning and associated whistler
waves exist at Venus. Some evidence of such waves was presented while
later analyses suggested that the spikes were largely produced by noise,
not waves {[}Taylor et al, 1987{]}. A more complete discussion of the
purported evidence of lightning and whistlers at Venus is given
elsewhere {[}Williams et al, 1983; Russell, 1991; Lorenz, 2018{]}. In
spite of all such discussions, it is the case that the electric field
instruments on these earlier satellites were not capable of fully
measuring dc and low-frequency electric fields because they operated in
a mode that made them more sensitive to current fluctuations than to dc
and low-frequency electric fields. The reason that the electric field
instruments were not in useful operating modes is that a working low
frequency electric field instrument requires application of a bias
current to the antennas in order that their net current from
photoemission, electron thermal flux, proton thermal flux, and other
minor current sources be close to zero, as is discussed in more detail
in the following section.

The Parker Solar Probe (PSP) flew biased electric-field detectors with
bias currents set for proper operation in sunlight. To provide Venus
gravity assists, PSP had seven near-Venus flybys, the first six of which
had this near-constant bias current. Four of these flybys (VGA1, VGA2,
VGA5, VGA6) encountered the foreshock, bow shock, or sunlit flanks, but
not the nightside magnetosphere, during which good electric field
measurements were made {[}George et al, 2024{]}. Two of the first six
flybys (VGA3 and VGA4) entered the nightside magnetosphere, where
because there was no sunlight, the current biasing of the antennas was
incorrect, and proper electric field measurements were not made. An
example of this situation occurred on VGA4 {[}George et al, 2023{]}, as
illustrated in Figure 1. In panel (1a), the spacecraft entered the
nightside Venusian magnetosphere at about 20:04:40 when the electric
field value changed from near zero to the unphysical -1000 mV/m, due to
the current imbalance from the loss of photoemission. The 50-150 Hz
bandpass filtered electric field in panel (1b) illustrates the purported
whistler wave that occurred at about 20:06:10. Because the detector was
more sensitive to current fluctuations than external potentials at this
time, this observation is more likely current noise and not a whistler
wave. This conclusion is reinforced in panel (1c) which provides the
hodogram of the filtered EX versus EY and shows that the structure was
linearly polarized and not at least partially circular polarized as
required of a whistler wave. This example illustrates the extreme
caution required when associating signals from improperly biased
electric field sensors with physical phenomena.

VGA-7, the last Venus flyby, occurred on November 6, 2024, and had a
bias current in the Venus shadow that allowed the nightside
magnetospheric electric field at Venus to be measured for the first
time. The relevant data for this measurement was provided by the Fields
instrument {[}Bale et al, 2017{]}. The biasing and the resulting fields
will be considered following a discussion of the algorithm for biasing the
electric field sensors.

\section{Biasing the electric field sensors}\label{biasing-the-electric-field-sensors}

The discussion of biasing electric field sensors requires consideration
of the simplified Langmuir probe theory. This theory gives the potential of
a body with respect to the nearby plasma by requiring that, in
equilibrium, the sum of all currents to the body be zero. Figure 2
provides a graphic illustration of this analysis with the potential of
the body relative to the nearby plasma described along the X-axis and
the current to the body given by the Y-axis. Suppose that the potential
of the body is positive. Then the entire electron thermal current is
attracted to become a negative current to the body, as illustrated in
the figure (neglecting focusing effects, secondary emission, etc., for
simplicity). When the body is negative, the electron thermal flux with
energies less than the potential of the body is reflected, so the
electron thermal current reaching the body decreases as the potential of
the body becomes more negative, with the current decreasing
exponentially with a scale equal to the electron temperature in a simple
model. With opposite signs, the same phenomena happen to the proton
thermal flux, with the proton current about a factor of 40 less than the
electron thermal current because of the heavier proton mass. In the
absence of photoemission or bias current (and neglecting other, smaller
currents), the voltage at which the two currents are equal in magnitude
and opposite in sign is at the illustrated negative voltage location
called the floating potential. At this location, a small change of
current produces a large change of the floating potential, as can be
seen in Figure 2, so the detector output is more dependent on current
than voltage fluctuations. To resolve this problem, the potential of the
body with respect to the nearby plasma should be nearly zero because, at
this voltage, a change of current produces a much smaller change of the
potential of the body. Thus, a positive bias current, such as that
illustrated in the figure, makes the sum of all the currents equal to
zero at the location labeled in the figure as the biased potential.

Applying the above discussion to an electric field detector, the
spacecraft is at the floating potential and each antenna should be
biased to be at or near the biased potential. The measured quantity,
called the spacecraft potential, is then the difference of the biased
potential and the floating potential which, for proper operation of the
detector, should be a few volts positive in the absence of sunlight, as
seen in the figure.

\section{Data}\label{data}

During VGA-7 on November 6, 2024, the spacecraft passed through the
nightside Venusian magnetosphere for about eight minutes at an altitude
as low as 400 kilometers. In the absence of photoemission, it was
desired to set the bias current to be about equal in magnitude to the
electron thermal current, as described in Figure 2. This requires
knowledge of the plasma density, whose average was well studied {[}Theis
et al, 1984{]}. However, the density at any altitude and solar-Venus
angle was shown to vary by as much as a factor of 10 depending on solar
wind conditions {[}Theis et al, 1984{]}. Rather than setting the bias
current according to the average density, it was decided to vary the
bias current by a factor of about three around the average value in
seven-second steps, in hope of achieving the correct bias at least some
of the time. The resulting VGA-7 bias current as a function of time is
illustrated in Figure (3a) with the negative values at times in sunlight
before entering the penumbra or after exiting it. The measured
spacecraft potential, in panel (3b) varied over a wide range with the
large negative values near the beginning and end of the crossing
signifying that the guessed plasma density was too small and the large
peaks at other times occurred because the guessed density was too large.
For a few minutes near the center of the crossing, the spacecraft
potential was small and positive, which is the requirement found from
the study of Figure 2 for a good electric field measurement. Panel (3c)
gives the measured electric field, which had a spike every seven seconds
due to the change of bias current and which showed unrealistically large
values when the spacecraft potential did not have the required small
positive value.

Figure 4 presents 30 minutes of data surrounding the closest approach to
Venus, which is illustrated in the spacecraft altitude plot in panel
(4a). The approximately eight-minute interval during which the
spacecraft was in the Venus shadow is defined by the loss of solar panel
current in panel (4b). One component of the electric field is given in
panel (4c), with the field in shadow appearing only when the spacecraft
potential of Figure (3b) had the appropriate small positive value. In
sunlight, the entire electric field is shown because the instrument was
properly biased for sunlight conditions and good wave measurements were
made {[}George et al, 2024{]}. Figures (4d) and (4e) give the spectra as
a function of time of the magnetic field and electric field, with spectra
for the electric field in shadow not illustrated because of the
difficulty in obtaining good spectra over such short time intervals. It
is noted that the electric field amplitude in shadow was similar to that
in sunlight, panel (4c), and that there was significantly less magnetic
field power in shadow than in sunlight, panel (4d). This suggests the
possible absence of low-frequency electromagnetic turbulence in the
nightside magnetosphere.

In Figures (5a) and (5b) the good electric field measurements in shadow
are given in single continuous plots. Their spectra, plotted in Figures
(5c) and (5d), show the presence of a few-Hz electric field wave with no
other features at frequencies below 1000 Hz. The average spectrum of the
electric field in panel (5e) confirms this result while the average
spectrum of the magnetic field in panel (5e) shows that no magnetic
field signal was observed below 1000 Hz other than 1/f noise (the
magnetic field data near 10 Hz was removed because of magnetic noise
from other sources at that frequency).

\section{Discussion}

Good electric field measurements were made in the nightside Venus
magnetosphere for the first time. During their one-minute observation
interval, the only wave mode observed was the few Hz electrostatic
noise. This result, plus the magnetic field measurements, showed that
there was little if any electromagnetic cascade turbulence, whistlers,
or other electromagnetic waves below 1000 Hz. To the extent that this
first one minute of good electric field measurements made in the
nightside magnetosphere can be generalized, the plasma environment of
Venus may be different from that of the terrestrial magnetosphere or the
solar wind and deserves further field and particle measurements to
understand the interaction between the solar wind and this unmagnetized
body.  The good electric field measurements signify the future approach towards routine electric field measurements in shadow when a biasing scheme appropriate to the plasma conditions is applied.

\backmatter

%

\bmhead{Acknowledgements}

This work was supported by NASA contract NNN06AA01C. The authors
acknowledge the extraordinary contributions of the Parker Solar Probe
spacecraft engineering team at the Applied Physics Laboratory at Johns
Hopkins University. The FIELDS experiment on the Parker Solar Probe was
designed and developed under NASA contract NNN06AA01C. Our sincere
thanks to M. Moncuquet, M. Pulupa, and P. Harvey for providing data
analysis material and for managing the spacecraft commanding and data
processing, which have become heavy loads thanks to the complexity of
the instruments and the orbit.

\bmhead{Open research}

The data used in this publication is available at
\href{http://spdf.gsfc.nasa.gov/}{spdf.gsfc.nasa.gov}, at
/data/spp/data/sci/fields/staging/l1b/dfb\_wf\_vdcX/2024, where X is 1,
or 2, or 3, or 4 for each of the four antennas.

\bmhead{Declarations}
Not applicable

\section{References}\label{references}
\mybibitem Bale, S. D., Goetz, K., Harvey, P. R., et al. 2016, SSRv, 204
\mybibitem Bridge, H. S., Lazarus, A. J., Snyder, C. W., Smith, E. J., Davis, Jr., L.,
\mybibitem Coleman, Jr., P. J., and Jones, D. E., 1967, Science, 158, 3809,
1669-1673, doi = 10.1126/science.158.3809.1669
\mybibitem George, H., Malaspina, D.M., Lee-Bellows, D., Gasque, L.C., Goodrich,
K., Ma, Y., 2024, Plasma wave survey from Parker Solar Probe
observations during Venus gravity assists, \emph{A\&A}, \textbf{689},
A214 \url{https://doi.org/10.1051/0004-6361/202450244}
\mybibitem George, H., Bale, S.D., Malaspina, D.M.,~Curry, S., Goodrich, K., Ma,
Y., Ramstad, R., and Conner, D., 2023, Non-Lightning-Generated Whistler
Waves in Near-Venus Space, \emph{Geophys. Res. Lett}., doi:
10.1029/2023GL105426
\mybibitem Gurnett, D. A., Kurth, W. S., Roux, A., et al. 1991, Science, 253, 1522
\mybibitem Gurnett, D. A., Zarka, P., Manning, R., et al. 2001, Nature, 409, 313
\mybibitem Lorenz, R. D. (2018). Lightning detection on Venus: A critical review.
Progress in Earth and Planetary Science, 5(1), 34.
\url{https://doi.org/10.1186/s40645-018-0181-x}
\mybibitem Russell, C. T. (1991). Venus lightning. In C. T. Russell (Ed.), (pp.
317--356). Springer Netherlands
\mybibitem Scarf, F. L., W.W.L. Taylor, C. T. Russell, L. H. Brace,~1980, Lightning
on Venus: Orbiter detection of whistler signals,~\emph{J. Geophys.
Res.},~\textbf{85},~8158--8166
\mybibitem Taylor, W.W.L., F. L. Scarf, C. T. Russell, L. H. Brace,~Evidence for
lightning on Venus,~\emph{Nature.},~\textbf{279},~614--616,~1979
\mybibitem Taylor, H. A. Jr., P. A. Cloutier, Z. Zheng,~Venus's lightning signals
reinterpreted as in situ plasma noise,~\emph{J. Geophys.
Res.},~\textbf{92},~9907--9919,~1987
\mybibitem Theis, R.F., Brace, L.H., Elphic, R.C., and Mayr, H.G., 1984, New
Empirical Models of the Electron Temperature and Density in the Venus
Ionosphere With Application to Transterminator Flow, JGR 89, 1477-1488
\mybibitem Williams, M. A., Krider, E. P., and Hunten, D. M. (1983). Planetary
lightning: Earth, Jupiter, and Venus. Reviews of Geophysics, 21(4),
892--902.https://doi.org/10.1029/RG021i004p00892

%
%
\newpage

\begin{figure}
\centering
\includegraphics[width=3.95909in,height=3.72222in]{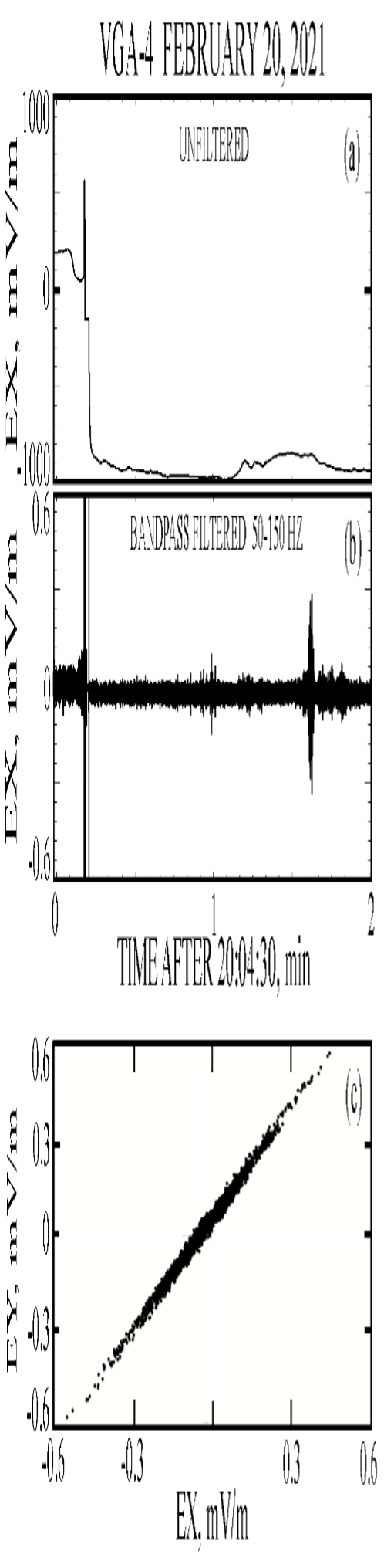}
\caption{
The electric field response when the spacecraft entered
shadow. Panel (1a) shows that the field decreased from a reasonable
value to the unphysical -1000 mV/m on entering the shadow. Panel (1b)
provides bandpass filtered data indicating a small signal in shadow that
is noise, and not a whistler wave, because the detector was near
saturation. Panel (1c) confirms this result because the wave is not
circular polarized, as would be required for a whistler.
}\label{fig1}
\end{figure}

\begin{figure}
\centering
\includegraphics[width=5.08264in,height=4.37183in]{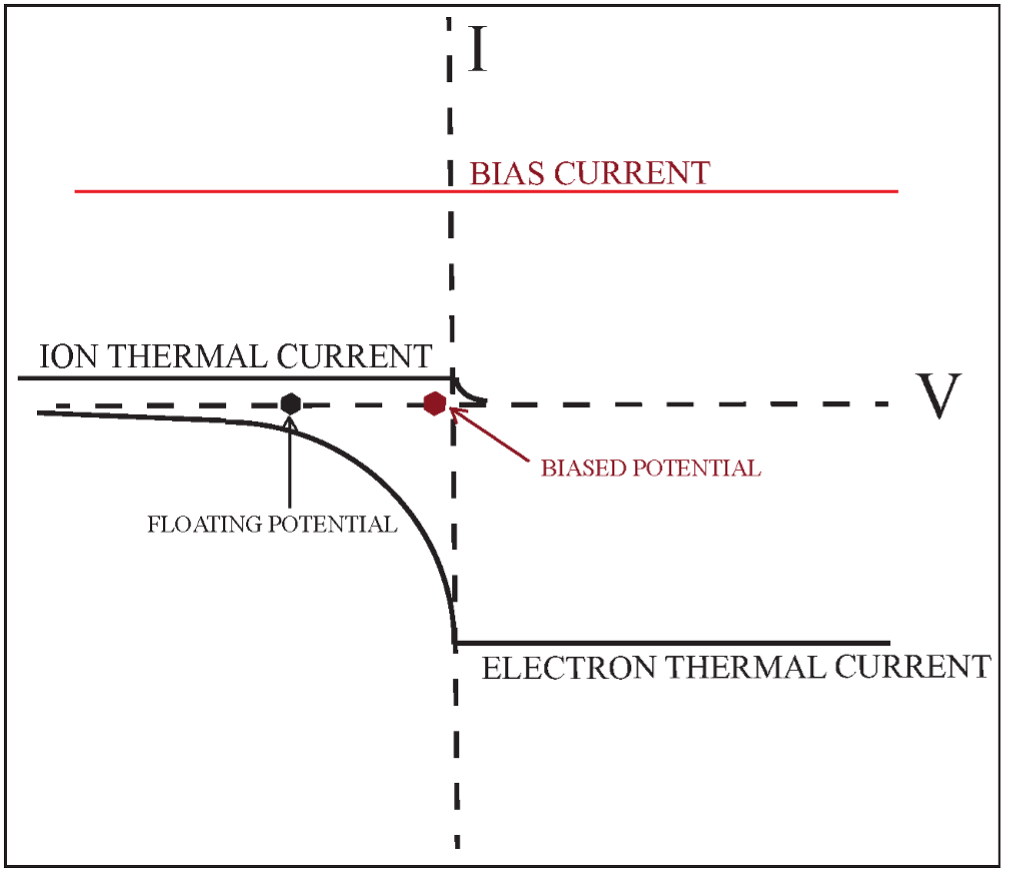}
\caption{Schematic of Langmuir probe theory showing the location of the
floating potential of a body experiencing only the proton and electron
thermal currents and the biased potential of a body that experiences the
additional bias current.}\label{fig2}
\end{figure}

\begin{figure}
\centering
\includegraphics[width=4.49306in,height=4.50505in]{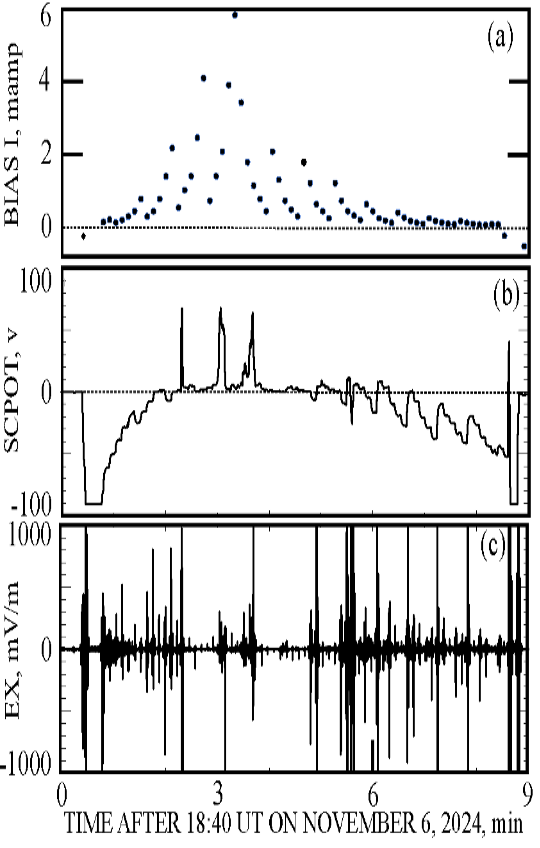}
    \caption{
Panel (3a) illustrates the bias current applied to the
antennas every seven seconds in an attempt to make the spacecraft
potential of panel (3b) a few volts positive. Panel (3c) gives the
electric field measured and that had a spike every seven seconds due to
the changing bias current and that was large when the spacecraft
potential differed from a small positive value.}\label{fig3}
\end{figure}

\begin{figure}
\centering
\includegraphics[width=5.12179in,height=4.12917in]{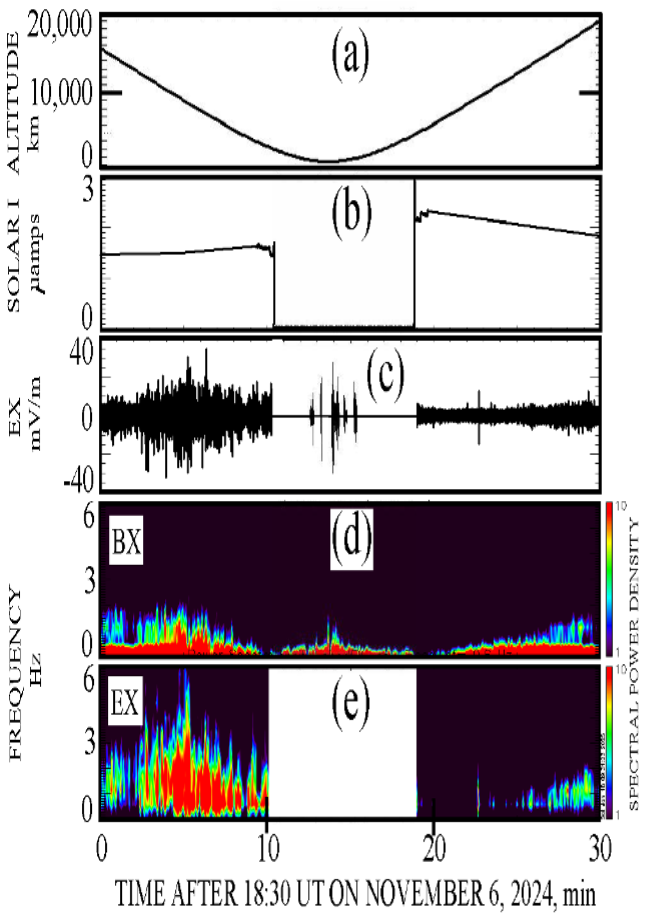}
\caption{
Panel (4a) gives the spacecraft altitude above Venus, which
was as small as 400 km. The approximately eight minute interval during
which the spacecraft was in the Venus shadow is defined by the loss of
solar panel current in panel (4b). One component of the electric field
is given in panel (4c), with the field in shadow appearing only when the
spacecraft potential had the appropriate small positive value. Figures
(4d) and (4e) give spectra of the magnetic field and electric field,
with that for the electric field in shadow not illustrated because of
the difficulty in obtaining good spectra over such short time intervals.
}
\label{fig4}
\end{figure}

\begin{figure}
\centering
\includegraphics[width=4.73108in,height=4.70833in]{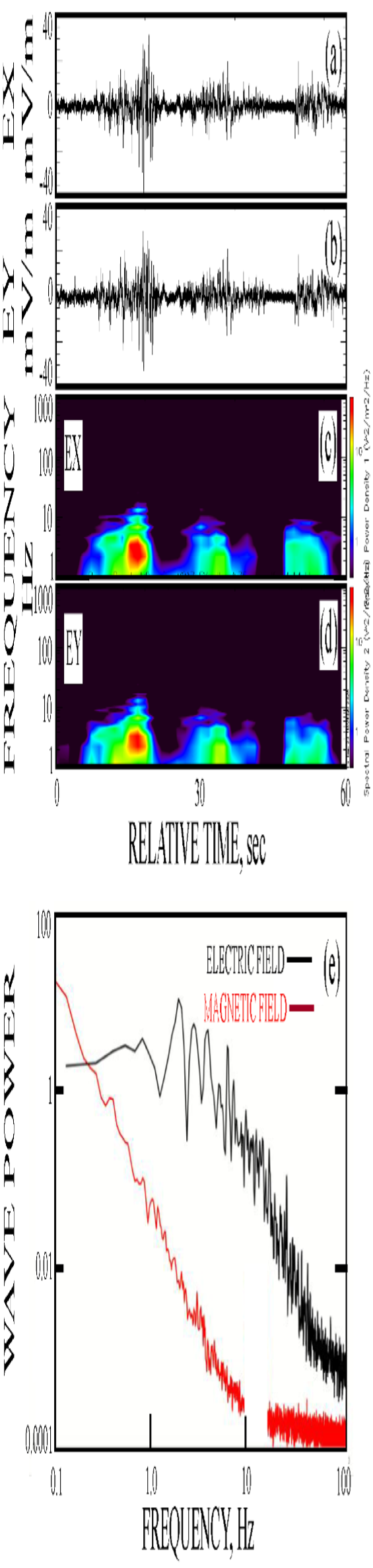}
\caption{Panels (5a) and (5b) give the good electric field measurements
in shadow as continuous plots. Their spectra, plotted in panels (5c) and
(5d), show the presence of a few Hz electric field wave with no other
features at frequencies below 1000 Hz. The average spectra of the
electric and magnetic field field in panel (5e) shows that no magnetic
field signal was observed below 1000 Hz.
}\label{fig5}
\end{figure}

\end{document}